%% file: main.tex
\documentclass[a4paper,10pt,aps,prd,showpacs,preprintnumbers,twocolumn,nofootinbib,superscriptaddress]{revtex4-2}

\usepackage[colorlinks=true,
            linkcolor=blue,
            citecolor=blue,
            urlcolor=blue]{hyperref}

\input{macros.tex}

\usepackage{amsfonts}
\usepackage{lettrine}
\usepackage{graphicx}
\usepackage{booktabs}
\usepackage{comment}
\usepackage{amsmath}
\usepackage{url}
\usepackage{gensymb}
\usepackage{soul}
\usepackage[utf8]{inputenc}
\usepackage{orcidlink}
\usepackage[normalem]{ulem}
\usepackage{empheq}

\def\dd{\mathrm{d}}

\begin{document}

\title{Luminosity-weighted galaxy and gravitational wave cross-correlations}

\author{Charles Dalang\orcidlink{0000-0002-7373-6903}}
\email{charles.dalang@ens.fr}

\affiliation{Institut Philippe Meyer, Département de Physique,\\ École Normale Supérieure (ENS), Université PSL, Paris, France
}

\author{Isabela Santiago de Matos\orcidlink{0000-0002-2686-2536}}
\email{isabela-matos@cbpf.ac.uk} 

\affiliation{Centro Brasileiro de Pesquisas Físicas, 22290-180, Rio de Janeiro, RJ, Brazil}

\affiliation{
Institute of Cosmology and Gravitation, University of Portsmouth, United Kingdom
}

\author{João Ferri\orcidlink{0009-0005-8123-2528} }
\email{joao.ferri@ipmu.jp}
\affiliation{Kavli IPMU (WPI), UTIAS, The University of Tokyo, Kashiwa, Chiba 277-8583, Japan}
\affiliation{Center for Frontier Science, Chiba University, 1-33 Yayoi-cho, Inage-ku, Chiba 263-8522, Japan}


\author{\\Raul\,Abramo\orcidlink{0000-0001-8295-7022}}
\email{raulabramo@usp.br}
\affiliation{Instituto de Física, Universidade de São Paulo,
Brazil}

\author{{Tessa\,Baker}\orcidlink{0000-0001-5470-7616}}
\email{tessa.baker@port.ac.uk}

\affiliation{
Institute of Cosmology and Gravitation, University of Portsmouth, United Kingdom
}


\date{\today }


\begin{abstract}
Cross-correlations of galaxies and gravitational waves (GW) can be used to extract cosmological parameters and the gravitational wave bias. In this work, we study how to include galaxy luminosities in these spatial cross-correlations, inspired by the standard approach of dark sirens line-of-sight redshift priors. We present a formalism to include luminosity information in galaxy-GW simulations and study the effects of assigning a hosting probability that depends on luminosity through a simple power law $p\e{host} \propto  L^\beta$ with $\beta \in \{0,1,2\}$ corresponding to uniformly, linear and quadratically weighted in luminosity. This effectively changes the GW linear bias with respect to dark matter and, in principle, allows to gain signal-to-noise and a sharper measurement of cosmological parameters. For a realistic luminosity dependence of the galaxy bias taken from DESI data, the gain is marginal, making the method robust to different luminosity weighting prescriptions. We show that the cross-correlation inference is able to recover the injected GW bias within $10\%$ if 
there is a mismatch between the injected linear bias of host galaxies and observed galaxies, provided that one sticks to relatively large scales. We also show that the Hubble constant and matter energy density can be recovered to $0.5\%$ and $5\%$ respectively with a third generation network of GW detectors, regardless of the modeling of the GW bias.
\end{abstract}

\maketitle

\section{Introduction} 

\lettrine{I}{}n the past decade, gravitational wave (GW) cosmology has evolved from a field of pure theoretical interest to one in which observations are quickly catching up \cite{Schutz:1986gp,Nissanke:2013fka,LIGOScientific:2026uyd}. The event GW170817 \cite{ LIGOScientific:2017zic} marked the first observation of a bright siren, which allowed for the first measurement of the Hubble constant using GWs \cite{LIGOScientific:2017adf}. The lack of alternative GW events with electromagnetic counterparts has led to the development of old and new ideas to infer cosmological parameters from the observation of GWs \cite{Schutz:1986gp,Ezquiaga:2022zkx, Gair:2022zsa}.
In particular, features in the mass spectrum of dark sirens can be used to infer cosmological parameters with a merger rate which is assumed to be isotropic and depend only on redshift \cite{Ezquiaga:2022zkx}. A galaxy catalogue can then help build a refined redshift line of sight prior to further constrain cosmological parameters \cite{Gair:2022zsa}. A major limitation of the latter method has turned out to be incompleteness in the galaxy catalogue as the quick rise of the importance of missing galaxies, which have been mostly assumed to be uniformly distributed in comoving volumes, wash out cosmological information (see \cite{Dalang:2023ehp,Dalang:2024gfk, Leyde:2024tov,Leyde:2025rzk} however for alternatives).

In parallel, cross-correlations of galaxies and GWs have emerged as a promising technique to naturally incorporate large-scale structure into GW cosmology \cite{Oguri:2016dgk, Mukherjee:2018ebj, Mukherjee:2020hyn, Bera:2020jhx, Diaz:2021pem, Scelfo:2022lsx, Mukherjee:2022afz, Fonseca:2023uay, Balaudo:2023klo, Zazzera:2024agl, Ferri:2024amc, Pedrotti:2025tfg,Sala:2025wwu, Pan:2025iya, SantiagodeMatos:2025iyj}, culminating in the \textit{Peak Sirens} method \cite{Ferri:2024amc} which has been applied to data for the first time in \cite{SantiagodeMatos:2025iyj}. A few major advantages stand out: it is relatively unsensitive to incompleteness \cite{Bera:2020jhx} and weak gravitational lensing \cite{Oguri:2016dgk, Ferri:2024amc}, it naturally incorporates large-scale structure \cite{Ferri:2024amc} and
its dependence on the mass spectrum appears to be sub-leading \cite{Cheng:2026atn}. This last point is crucial since it requires many extra parameters (up to 25 in current LIGO-Virgo-Kagra (LVK) analysis \cite{LIGOScientific:2025jau}), which degrades the cosmological constraints. 

Luminosity weighting of galaxies can make a substantial difference to GW events which are well localized in the dark sirens method \cite{Gray:2019ksv,Finke:2021aom}. Indeed, this allows to boost the signal of observed galaxies with respect to the unobserved fainter ones. The purpose of this work is to clarify the role of luminosity weighting in the cross-correlation method and how to implement it in practice in the \textit{Peak Sirens} method \cite{Ferri:2024amc}. We show how the signal to noise ratio of the cross-correlation is expected to be proportional to the GW bias, which is inherited from the GW host galaxies. We run simulations to study the impact of luminosity weighting of galaxies on the uncertainties of cosmological parameters and on the GW bias, and apply the formalism to a galaxy catalogue that has similar properties to the DESI DR1 Bright Galaxy Survey. We use different luminosity weighting models and show that the method is robust to these changes in the weighting prescriptions. We further show that the method is able to recover the GW bias if the GW hosts are simulated to come from a population of galaxies with a different bias than the galaxies that are observed.

This work is structured as follows. In Sec.\,\ref{sec:Methodology}, we describe the methodology that we use for this analysis and fix the notation. In Sec.\,\ref{sec:Galaxy_GW_simulations}, we use DESI DR1 as an inspiration for luminosity-weighted galaxy simulations and describe how that enters in the Peak sirens simulations. In Sec.\,\ref{sec:results}, we apply the methodology to a DESI-like Bright Galaxy survey with third generation GW simulated data, discuss our results and conclude in Sec.\,\ref{sec:Conclusion}.

\section{Methodology} \label{sec:Methodology}
 
Galaxies with different luminosities can cluster with a different linear biases, although other factors may count, such as color or metallicity. As we will see, the signal to noise ratio of the cross-correlation coefficients is proportional to the gravitational wave bias, which is inherited from the GW host galaxies. Therefore, it is crucial to correctly determine the clustering properties of the host galaxies. The ability of a galaxy to host a GW merger may depend on many things, including stellar mass, metallicity and age. The GW bias value is likely to encode the physics of the formation channels of compact binaries and their environments. It could even be affected by the time it takes for a binary to merge \cite{Dehghani:2024wsh}. In the following, we consider luminosity as an indicator of how likely a galaxy is to host a GW merger. We outline how this information may be used in the cross-correlation of galaxies and gravitational waves to extract cosmological parameters. 

\subsection{Luminosity-dependent catalogue}

We take a galaxy catalogue, which has the following information; right ascension $\alpha$ and declination $\delta$ which define a unit vector on the sky $\bs{\hat{n}}$, spectroscopic redshift $z$ and apparent magnitude $m$ in a certain band. Given a fiducial cosmology model, the luminosity $L$ of a galaxy can be computed from its apparent magnitude and redshift as
\begin{align}
\log_{10}\l(\frac{L}{L_0}\r) = -\frac{2}{5} \l[ m - 5 \log_{10}(D_L(z)) -25 - k(z)\r]\,,\label{eq:Luminosity}
\end{align}
where $k(z)$ indicates a $k$-correction \cite{Hogg:2002yh} and $D_L(z)$ is the luminosity distance calculated in the fiducial cosmology. We can construct the density contrast of the luminosity-dependent galaxy map by subtracting and normalizing by the monopole
\begin{align}
\delta_L (z, \bs{\hat{n}}) & = \frac{n\e{g}(z, \bs{\hat{n}}, L) - \bar{n}\e{g}(z, L)}{\bar{n}_{\rm{g}}(z, L)} \,, \label{eq:density_contrasts}
\end{align}
where $n\e{g}(z,\bs{\hat{n}},L)$ is short for the number density of galaxies $N\e{g}$ with respect to its arguments, such as
\begin{align}
n\e{g}(z,\bs{\hat{n}},L) \equiv \frac{\dd N_{\rm{g}}(z, \bs{\hat{n}}, L) }{\dd z \dd \Omega \dd L}\,.
\end{align}
The monopole of galaxies, denoted with a bar, can be extracted from a full-sky map as
\begin{align}
\bar{n}\e{g}(z,L) = \frac{1}{4\pi} \int_{\mathbb{S}_2} \dd \bs{\hat{n}} \, n_{\rm{g}}(z, \bs{\hat{n}}, L)\,.\label{eq:monopole}
\end{align}
In practice, there may be subtleties in extracting the monopole due to observational challenges which artificially affect the number density of sources as a function of direction on the sky that we do not consider in this work. These maps can be decomposed into spherical harmonics
\begin{align}
\delta_L (z, \bs{\hat{n}}) & = \sum_{\ell=0}^{\ell\e{max}} \sum_{m=-\ell}^\ell a^{L}_{\ell m}(z) Y_{\ell m}(\bs{\hat{n}}) \,,
\end{align}
thereby defining harmonic coefficients which carry a luminosity bin index $L$. Statistical isotropy enforces
\begin{align}
\langle a^{L}_{\ell m}  (a^{L}_{\ell' m'})^*\rangle & = C_\ell^{L}(z) \delta_{\ell \ell'}\delta_{m m'}\,, \label{eq:C^L_ell}
\end{align}
which defines the auto-correlation $C_\ell^{L}(z)$ for a certain luminosity bin $L$. One can estimate the auto-spectrum of galaxies in a certain luminosity bin with the following quadratic estimator
\begin{align}
\hat{C}^{L}_\ell(z_i) = \frac{1}{2\ell +1} \sum_{m=-\ell}^\ell a^{L}_{\ell m}(z_i)[a_{\ell m}^{L}(z_i)]^*\,.
\end{align}
On linear scales, the density contrast of a certain luminosity bin is proportional to the density contrast of dark matter, i.e.\,$\delta_L(z, \bs{\hat{n}}) = b_L(z)\delta_{\rm{dm}}(z,\bs{\hat{n}})$, where the bias $b_L$ only depends on redshift. This allows to define the square of the luminosity-dependent bias $b_L$, which corresponds to the ratio of the power spectra of galaxies in a certain luminosity bin with respect to dark matter, as
\begin{align}
b_L^2(z) = C^{L}_\ell(z)/C^{\rm{dm}}_\ell(z)\,.\label{eq:bias_L}
\end{align}
When performing cross-correlations of galaxies and GWs, one can integrate the number density of galaxies over luminosity with a certain weight $\beta$
\begin{align}
n\e{g}(z,\bs{\hat{n}}) = \int_{L\e{min}}^{L\e{max}} \dd L \,L^\beta \, n\e{g}(z,\bs{\hat{n}},L)\,,\label{eq:weighted_counts}
\end{align}
where $\beta =0, 1, 2$ for uniform, linear and quadratic luminosity weighting, respectively. Note that the weight does not necessarily need to be an integer. The standard procedure in cross-correlations is to work with uniform weighting of galaxies ($\beta=0$). In the dark sirens method, it is customary to use weights $\beta = 0$ and $1$ \cite{LIGOScientific:2026uyd}.

The galaxy catalogue defined by Eq.\,\eqref{eq:weighted_counts} inherits an effective average galaxy bias, which is luminosity-weighted with weight $\beta$
\begin{align}
b\e{eff}(z) = \frac{\sum_{L=1}^{N_L} L_L^\beta b_L(z) }{\sum_{L=1}^{N_L} L_L^\beta}\,. \label{eq:effective_bias} 
\end{align}
The crucial idea is that if one weights the galaxies with their luminosities according to $L^\beta$ to construct a luminosity-weighted galaxy map to sample the Binary Black Hole (BBH) host galaxies, then the BBH inherit the effective bias defined in Eq.\,\eqref{eq:effective_bias}. This may boost the cross-correlation signal, as we will see.

\subsection{Auto- and cross-spectra}

One can define the luminosity-weighted galaxy density contrast $\delta\e{g}(z,\bs{\hat{n}})$ and GW density contrast $\delta\e{gw}(D_L,\bs{\hat{n}})$ in a way which is analogous to Eq.\,\eqref{eq:density_contrasts}, using the luminosity weighted counts (Eq.\,\eqref{eq:weighted_counts}). That is
\begin{align}
\delta\e{g} (z, \bs{\hat{n}}) & = \frac{\l( n_{\rm{g}}(z, \bs{\hat{n}}) -  \bar{n}_{\rm{g}}(z) \r)}{\bar{n}_{\rm{g}}(z)} \,,\\ 
\delta\e{gw} (D_L, \bs{\hat{n}}) & = \frac{\l( n_{\rm{gw}}(D_L, \bs{\hat{n}}) - \bar{n}_{\rm{gw}}(D_L)\r)}{\bar{n}_{\rm{gw}}(D_L)} \,, 
\end{align}
The monopoles are defined as in Eq.\,\eqref{eq:monopole}. In the following, we consider GWs emitted from BBH for simplicty. \footnote{In principle, nothing prevents from including binary neutron stars or neutron star-black hole systems in the GW skymaps.} A key aspect of the Peak Sirens method is the fact that the two contrast maps, which trace the same underlying matter field, are observed in different radial coordinates -- redshift for galaxies and luminosity distances for GWs. Since the correspondence between these coordinates is cosmology-dependent, maximizing the correlations between those maps allows inference of the $D_L(z)$ relation and the background cosmological parameters in it, analogously to the Alcock-Paczynski test \cite{1979Natur.281..358A}.

One may expand the density contrasts in spherical harmonics, which depend either on $z$ or $D_L$,
\begin{align}
\delta\e{g} (z, \bs{\hat{n}}) & = \sum_{\ell=0}^{\ell\e{max}} \sum_{m=-\ell}^\ell a^{\rm{g}}_{\ell m}(z) Y_{\ell m}(\bs{\hat{n}}) \,,\\
\delta\e{gw} (D_L, \bs{\hat{n}})  & = \sum_{\ell=0}^{\ell\e{max}} \sum_{m=-\ell}^\ell a^{\rm{gw}}_{\ell m}(D_L) Y_{\ell m}(\bs{\hat{n}})\,.
\end{align}
In a statistically isotropic Universe, the auto and cross-spectra satisfy
\begin{align}
\langle a^{\rm{g}}_{\ell m}  (a^{\rm{g}}_{\ell' m'})^*\rangle & = C_\ell^{\rm{g}}(z) \delta_{\ell \ell'}\delta_{m m'}\,,\\
\langle a^{\rm{gw}}_{\ell m}(a^{\rm{gw}}_{\ell' m'})^*\rangle & = C_\ell^{\rm{gw}}(D_L) \delta_{\ell \ell'}\delta_{m m'}\,,\\
\langle a^{\rm{g}}_{\ell m} (a^{\rm{gw}}_{\ell' m'})^*\rangle & = C_\ell^{\rm{g gw}}(z,D_L) \delta_{\ell \ell'}\delta_{m m'}\,.
\end{align}
Note that in contrast to Eq.\,\eqref{eq:C^L_ell}, the galaxy auto-spectra does not carry a luminosity index, since the galaxy count map has been weighted by luminosity over all luminosity bins, according to Eq.\,\eqref{eq:weighted_counts}. The auto-spectrum of galaxies is an important cosmological probe of the growth of structures. The cross-spectrum can directly be used to constrain the Hubble diagram, while the auto-correlation of GWs is weakly informative, as it is dominated by shotnoise in the setting that we consider. Implementing luminosity weighting amounts to sample the host BBH with 
in a way that depends on luminosity, as
\begin{align}
p\e{host}(L) \propto L^\beta\,,
\end{align}
with $\beta = 0,1,2$ for uniform, linear and quadratic luminosity weighting, respectively. 

In practice, to extract cosmological parameters from the cross-correlations $C_\ell^{\rm{g gw}}(z,D_L)$, one requires to build simulations which mimic the data as closely as possible. Here, we point out that if one wants to include some form of luminosity weighting in the simulations, one should include a luminosity-dependent bias for each population of galaxies. This way, one can build sky maps for each population, which respect their clustering properties. Including the luminosity weighting then amounts to weighting each of these maps proportionally to $p\e{host}(L)$, as in Eq.\,\eqref{eq:weighted_counts}. 

To estimate the $C_\ell(z_i, D_{Lj})$'s in each bin from full sky information, one can use the following quadratic estimator
\begin{align}
\hat{C}_\ell^{\rm{g}\rm{gw}}(z_i,D_{Lj}) = \frac{1}{2\ell +1} \sum_{m=-\ell}^\ell a_{\ell m}^{\rm{g}}(z_{i})(a_{\ell m}^{\rm{gw}}(D_{Lj}))^*\,.
\end{align}
We use a Gaussian likelihood
\begin{align}
\log\mathcal{L} = -\frac{1}{2}\sum_{\ell i j, \ell' i' j'} \Delta C_\ell^{ij} [\hbox{Cov}^{-1}]_{\ell i j, \ell' i' j'} \Delta C_{\ell'}^{i'j'}\,,
\end{align}
where $\Delta C_\ell^{ij}$ is the difference between the observed and theoretical cross-correlations, i.e.\,$\Delta C_\ell^{ij} = C^{\rm{obs}}_\ell(z_i, D_{Lj}) - C^{\rm{th}}_\ell(z_i, D_{Lj})$. The fiducial cosmology theoretical $C_\ell^{\rm{th}}$ are estimated from 
the mean spectrum of 1000 simulations, built with a given prescription of $p\e{host}$, and then further multiplied by a free GW bias parameter $b\e{gw}$, that we seek to determine jointly with cosmological parameters. This implies that we are determining the GW bias with respect to the luminosity-weighted galaxies. Explicitly, we define the GW bias as\footnote{An equivalent definition is $b\e{gw}^2 \equiv C_\ell^{\rm{gw}}/C_\ell^{\rm{g}}$. However, the autospectrum of GWs is noisier than the cross-spectrum due to shot noise. Hence, we privilege the former definition.}
\begin{align}
b\e{gw} \equiv C_\ell^{\rm{g gw}}/C_\ell^{\rm{g}}\,.\label{eq:gw_bias}
\end{align}
Note that it is defined here with respect to the luminosity weighted galaxies and not directly with respect to the dark matter power spectrum. The theoretical $C_\ell^{\rm{th}}$ for a different cosmological model are computed using a rescaling technique introduced in \cite{Ferri:2024amc} and applied to data in \cite{SantiagodeMatos:2025iyj}. Since we present a forecast, we use as the observed $C_\ell^{ij}$ the mean of the simulated $C_\ell^{ij}$ in the fiducial cosmology, although it would also be fine to use one realization of the simulations. We estimate the inverse covariance, using the Gaussian approximation \cite{Abramo:2021irg,Abramo:2022qir}
\begin{align}
& [\hbox{Cov}^{-1}]_{\ell i j, \ell' i' j'}  = \frac{2\ell +1}{4} (2-\delta_{ij})(2-\delta_{i' j'}) \nonumber \\
& \quad \times \l( (C_\ell^{-1})^{ii'} (C_\ell^{-1})^{jj'} + (C_\ell^{-1})^{ij'} (C_\ell^{-1})^{i'j} \r)\,.  \label{eq:covariance}
\end{align}
Note that in some cases, sample covariance is a more accurate estimate, specially when the number of GW events is low, as in \cite{SantiagodeMatos:2025iyj}. 

\subsection{Signal to noise ratio}
In this section, we estimate the signal to noise ratio (SNR) of each mode and show that it scales with the GW bias and host galaxy bias. The SNR is defined as 
\begin{align}
\hbox{SNR}^{ij}_\ell & \equiv \frac{(C_\ell^{\rm{ggw}})^{ij}}{\sqrt{(\hbox{Cov})^{ij,ij}_\ell}}\\
& = \frac{(2\ell + 1) (C_\ell^{\rm{ggw}})^{ij}}{\l((C_\ell^{\rm{g}})^{ii} (C_\ell^{\rm{gw}})^{jj} + ((C_\ell^{\rm{ggw}})^{ij})^2\r)^{1/2}}\,.\label{eq:SNR}
\end{align}
We consider a situation in which galaxies are abundant while GWs are sparse. In that case, the GW auto-spectrum is dominated by shot noise, while the galaxy shot noise is subdominant on large scales. In the linear regime, the biases can be factored out as in Eq.\,\eqref{eq:bias_L}, to find
\begin{align}
(C_\ell^{\rm{g}})^{ii} & \simeq (b\e{g}^i)^2 (C_\ell^{\rm{dm}})^{ii} + \frac{1}{4 \pi N\e{g}^i} \,,\\
(C_\ell^{\rm{gw}})^{jj} & \simeq \frac{1}{4 \pi N\e{gw}^j}\,,\\
(C_\ell^{\rm{ggw}})^{ij} & \simeq (b\e{g}^i)^2 b\e{gw}^j
(C_\ell^{\rm{dm}})^{ij}\,.
\end{align}
We recall that we defined the GW bias with respect to the galaxy autospectrum and not directly with respect to the dark matter autospectrum, as in Eq.\,\eqref{eq:gw_bias}. For the sake of this argument, we are neglecting the effect of the GW large localization and distance errors in its shot noise. Plugging this in the expression for the SNR \eqref{eq:SNR}, we obtain
\begin{align}
\hbox{SNR}^{ij}_\ell & \simeq \frac{(2\ell + 1)(b^i\e{g})^2 b\e{gw}^j (C_\ell^{\rm{dm}})^{ij}}{\sqrt{\frac{1}{4\pi N\e{gw}^j} \l( (C_\ell^{\rm{dm}})^{ii} + \frac{1}{4 \pi N\e{g}^i (b\e{g}^i)^2} \r) + (b\e{gw}^j ((C_\ell^{\rm{g}})^{ij})^2 }}
\end{align}
For a sufficiently low number of GWs in bin $j$ and sufficiently many galaxies in bin $i$, the first term dominates the last two in the denominator and we are left with 
\begin{align}
\hbox{SNR}^{ij}_\ell & \simeq (2\ell + 1)(b^i\e{g})^2 b\e{gw}^j  \sqrt{ \frac{4\pi 
N\e{gw}^j ((C_\ell^{\rm{dm}})^{ij})^2}{(C_\ell^{\rm{dm}})^{ii} }}\,. \label{eq:SNR_bgw}
\end{align}
Notably, the SNR scales with the bias combination $(b^i\e{g})^2 b^j\e{gw}$ which corresponds to the bias of GWs with respect to dark matter. The GWs inherit the bias of the GW host galaxies and can be further biased with respect to those. In the following, we explore different weighting prescriptions for galaxies of varying intrinsic luminosity. It is expected that more luminous galaxies should host a higher proportion of BBHs. Those are also expected to be more clustered than fainter galaxies, thereby having a higher bias which in turn, affects the GW bias and the SNR via Eq.\,\eqref{eq:SNR_bgw}. In the following sections, we explore the bias of galaxies as a function of luminosity and assess the potential impact on the cross-correlations.

\section{Galaxy and gravitational wave simulations}\label{sec:Galaxy_GW_simulations}

The goal here is to take a realistic input for simulations which are supposed to mimic DESI DR3-like data and to outline how this affects the Peak Sirens simulations.

\subsection{DESI-like Bright Galaxy survey simulations}

We use DESI DR1 bright galaxy data to infer the luminosity-dependent biases and number density of galaxies. We then assume that the same number density is observed over a DESI DR3-like footprint and use that as input for our forecasts. Rather than focusing on reproducing exactly the properties of DESI DR3, the goal is to compare different luminosity weighting schemes for GW-cosmology with cross-correlations with galaxies with a typical stage-IV spectroscopic survey.

\begin{figure}
    \centering
    \includegraphics[width=1.0\linewidth]{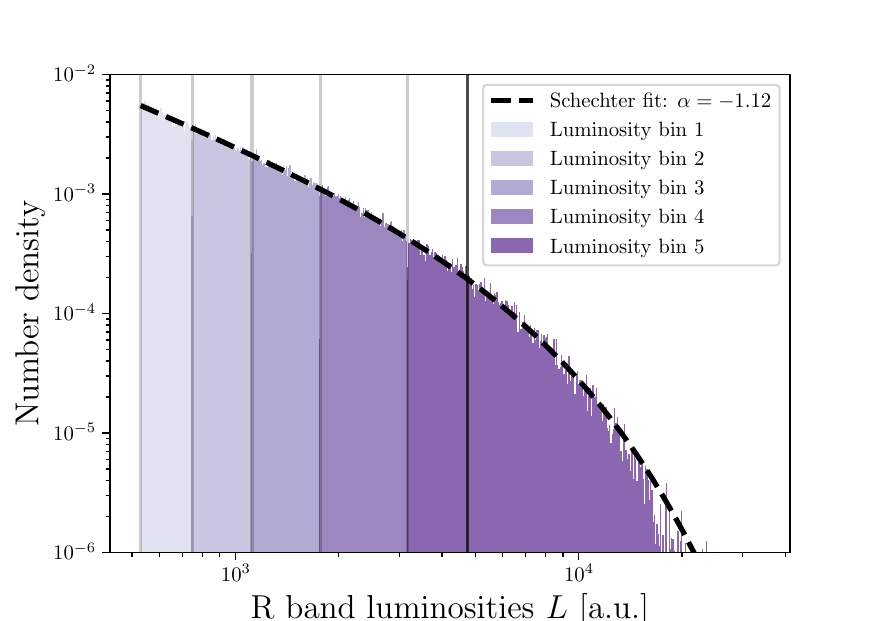}
    \caption{Here, we show the luminosity distribution together with the 5 luminosity bins. It obeys a Schechter function \cite{Schechter:1976iz} with a decaying power law at low luminosities and an exponential cutoff at $L_*$ indicated by the vertical black line.}
    \label{fig:Luminosity_bins_5Lbins}
\end{figure}
\begin{figure}
    \centering
    \includegraphics[width=1.0\linewidth]{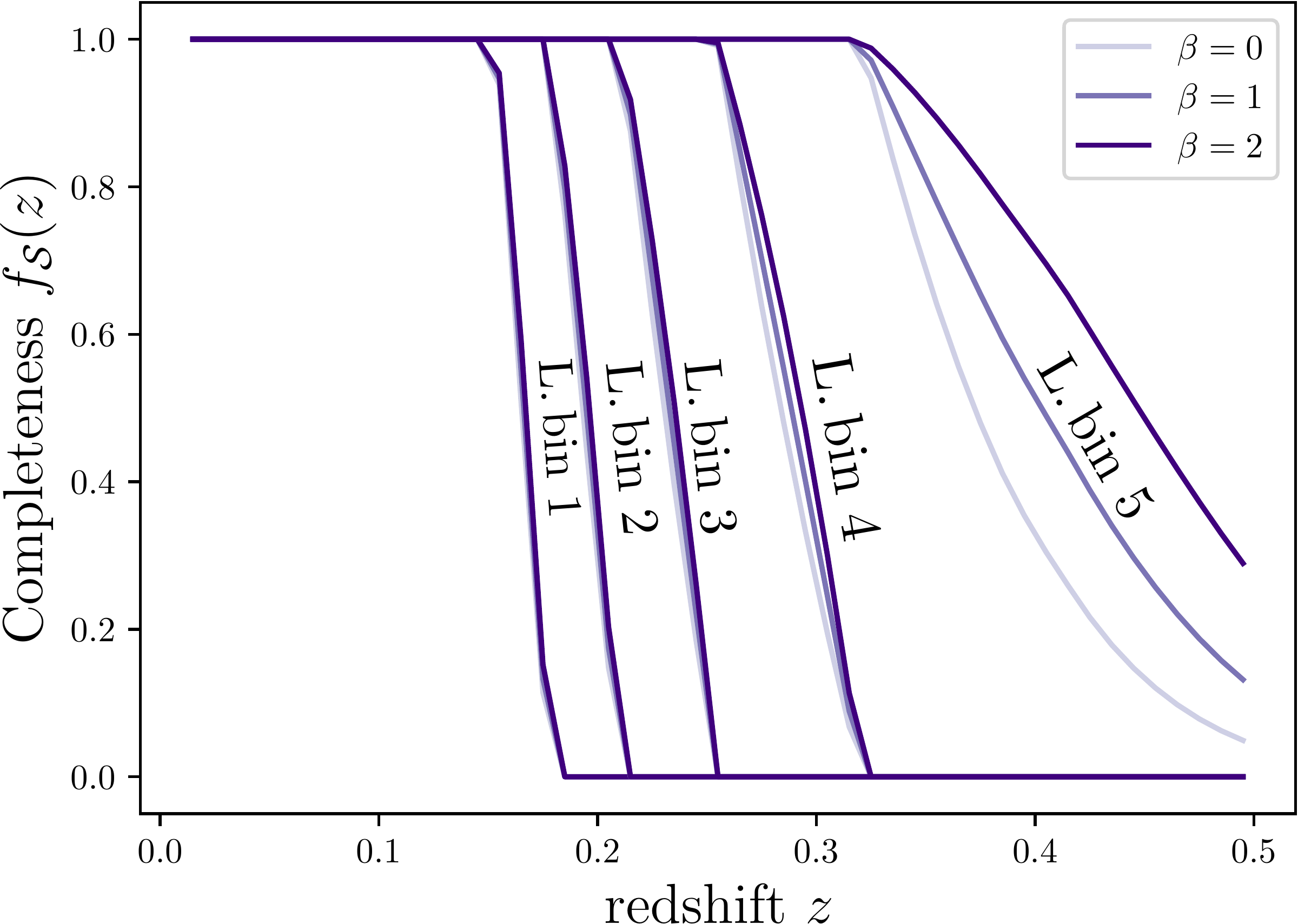}
    \caption{We present the completeness as a function of redshift for the 5 luminosity bins and 3 different luminosity weighting prescriptions. At redshift $z\leq 0.1$, all luminosity bins are complete. The faintest galaxies drop to zero in completeness at the lowest redshift. The higher luminosity bin galaxies also drop in completeness to zero at higher redshifts, as expected. The luminosity weighting plays a small role in the completeness.
    }
    \label{fig:Completeness}
\end{figure}
\begin{figure}
    \centering
    \includegraphics[width=0.8\linewidth]{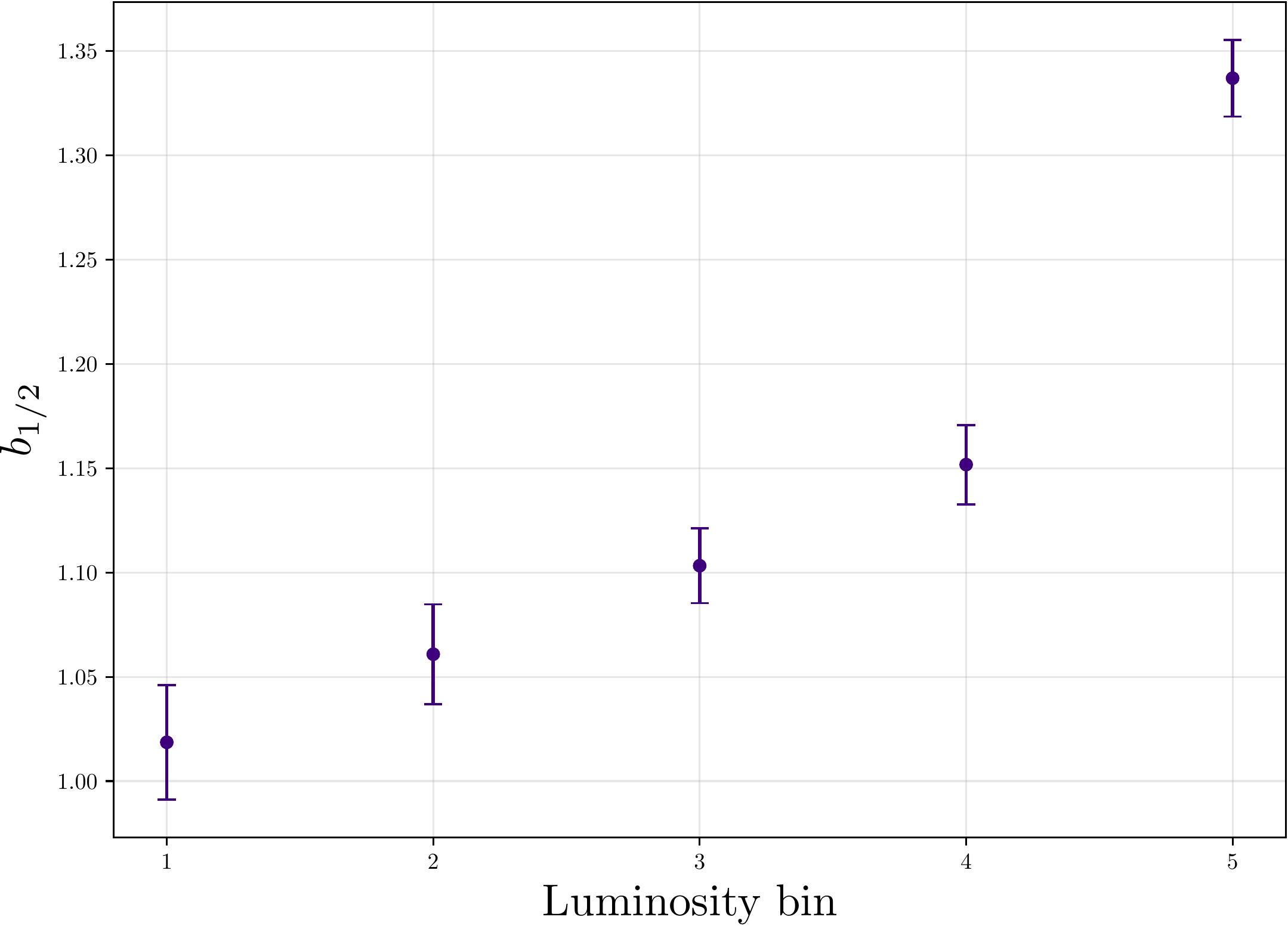}
    \caption{Here, we show the galaxy bias model parameter $b_L(z) = b_{1/2} (1+z)^{1/2}$ for the 5 bins  corresponding to the equipopulated luminosity bins. In terms of luminosity, these represent $[5.6, 8.2, 12.6,21.2, 52.2] \%$ of the total luminosity. }
    \label{fig:bias_parameter_L_NLbins=5}
\end{figure}

We take DESI DR1 data for bright galaxies. From Eq.\,\eqref{eq:Luminosity}, we estimate the luminosity of galaxies in the fiducial cosmology. We use {\tt kcorrect} \cite{2007:Blanton} to estimate the $k$-correction as a function of redshift $k(z)$ from a subsample of galaxies using information in the $g-$, $r-$ and $z-$ bands. We plot a histogram of luminosities in Fig.\,\ref{fig:Luminosity_bins_5Lbins}. The Bright Galaxies are split into $5$ equi-populated luminosity bins, which defines luminosity bin edges $L\e{edge}^i$. The 
mean luminosity of each bin is computed as
\begin{align}
L\e{bin}^i = \frac{\int_{L\e{edge}^i}^{L\e{edge}^{i+1}} \dd L L \phi(L)}{\int_{L\e{edge}^i}^{L\e{edge}^{i+1}} \dd L \phi(L)}
\end{align}
where $\phi(L)$ is the luminosity function of the bright galaxies. This defines the weight that we apply to each number count map. We fit a Schechter function \cite{Schechter:1976iz} to the luminosity distribution shown in Fig.\,\ref{fig:Luminosity_bins_5Lbins} 
\begin{align}
\phi(L) = \phi_* \l( \frac{L}{L_*}\r)^{-\alpha} \hbox{exp}(-L/L_*)\,,
\end{align}
and find $L_*/L_0 = (4.7 \pm 0.2)\times 10^{3} $ and $\alpha = -1.12\pm 0.02$. We compute the BBH hosting probability as
\begin{align}
p\e{host}(L\e{bin}^i) = \frac{(L\e{bin}^i)^\beta}{ \sum_i^{N_L} (L\e{bin}^i)^\beta}\,.
\end{align}
\begin{figure}
    \centering
    \includegraphics[width=1.0\linewidth]{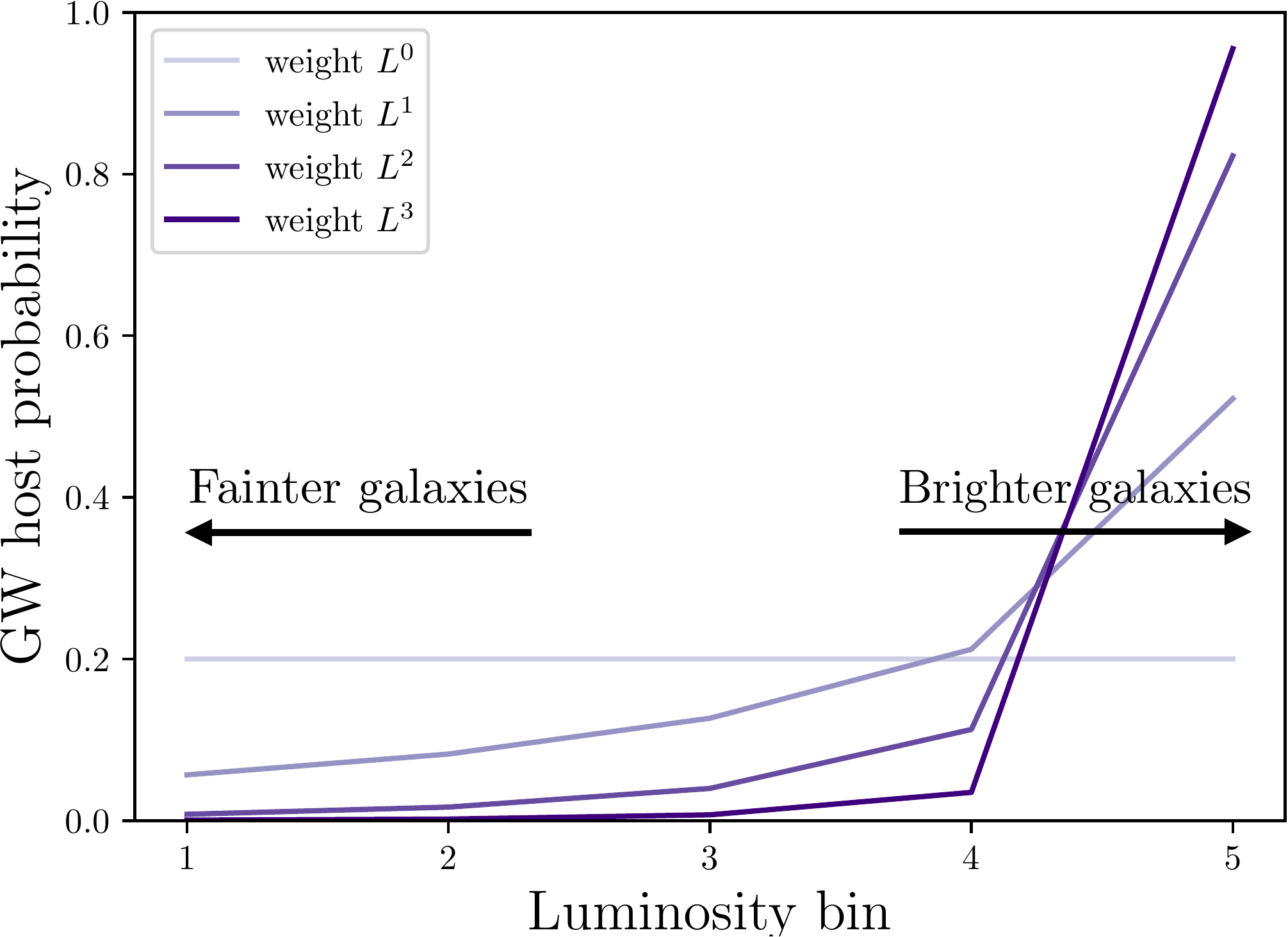}
    \caption{Here, we show the probability of hosting a GW for each of the 5 luminosity bins depending on the power $\beta$ in  $p\e{host}(L) \propto L^\beta$.}
    \label{fig:Lweight}
\end{figure}
We plot the resulting probabilities in Fig.\,\ref{fig:Lweight} for different weights $\beta$. We also define a completeness map as 
\begin{align}
f^i_\mathcal{S}(z) = \frac{\int_{L\e{th}^i(z)}^{L\e{edge}^{i+1}} \dd L L^\beta \phi(L)}{\int_{L\e{edge}^i}^{L\e{edge}^{i+1}} \dd L \phi(L)}
\end{align}
where 
\begin{align}
L^i\e{th}(z) = \max(L\e{th}(z,m\e{th}),L\e{edge}^i) \Theta(L\e{edge}^{i+1}-L\e{th}(z,m\e{th}))
\end{align}
and $L\e{th}(z,m\e{th})$ is calculated from the magnitude threshold $m\e{th} =m\e{th}(\bs{\hat{n}})$, using Eq.\,\eqref{eq:Luminosity} and $\Theta(x)$ indicates the Heaviside step function. For observations, one would typically estimate magnitude threshold pixel by pixel following Ref.\,\cite{Gray:2021sew}. For simplicity and for this forecast, we take $m\e{th}= \max_{\bs{\hat{n}}} (m\e{th}(\bs{\hat{n}}))$, which mimicks an optimistic survey where the completeness is as good in all pixels as in the best observed pixel of DESI DR1. We present the completeness of the 5 luminosity bins in Fig.\,\ref{fig:Completeness} for different luminosity weighting prescriptions.
In order to build galaxy simulations with luminosity information, we compute the bias in different luminosity bins as per Eq.\,\eqref{eq:bias_L}, using the luminosity-dependent maps. We parametrize their redshift dependence as $b_L(z) = b_{1/2}(1+z)^{1/2}$ and fit for the parameter $b_{1/2}$, which we present in Fig.\,\ref{fig:bias_parameter_L_NLbins=5} for the five different luminosity bins. This allows to have an estimate of the bias as a function of redshift even in regions where the completeness drops to zero. One can see a clear monotonically increasing trend as the luminosity increases. Brighter galaxies are more clustered than fainter ones. We are then ready to proceed and run the simulations with the different weightings of galaxies. In particular, we can consider different hosting probability weights for these populations of galaxies, as shown in Fig.\,\ref{fig:Lweight}. 

\begin{figure}
    \centering
    \includegraphics[width=1.0\linewidth]{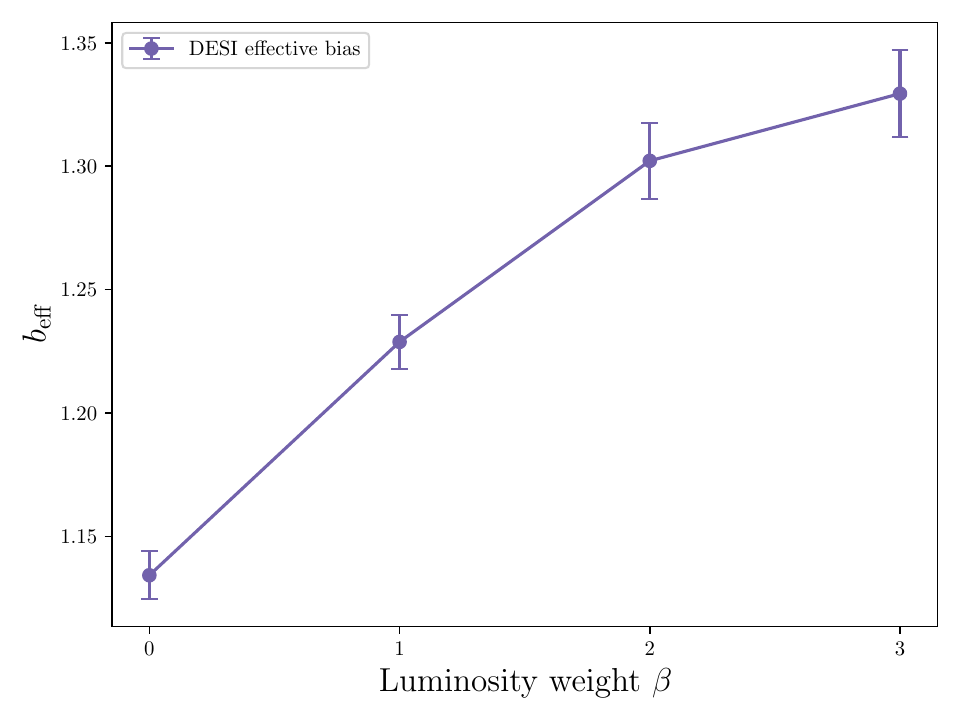}
    \caption{Here, we show the effective bias for different weights $\beta$. As one increases the weight $\beta$, the higher luminosity galaxies which have a higher bias weight more than their fainter counter part, increasing the effective bias from Eq.\,\eqref{eq:effective_bias}. }
    \label{fig:Effective_bias}
\end{figure}

\begin{figure}
    \includegraphics[width=1.0\linewidth]{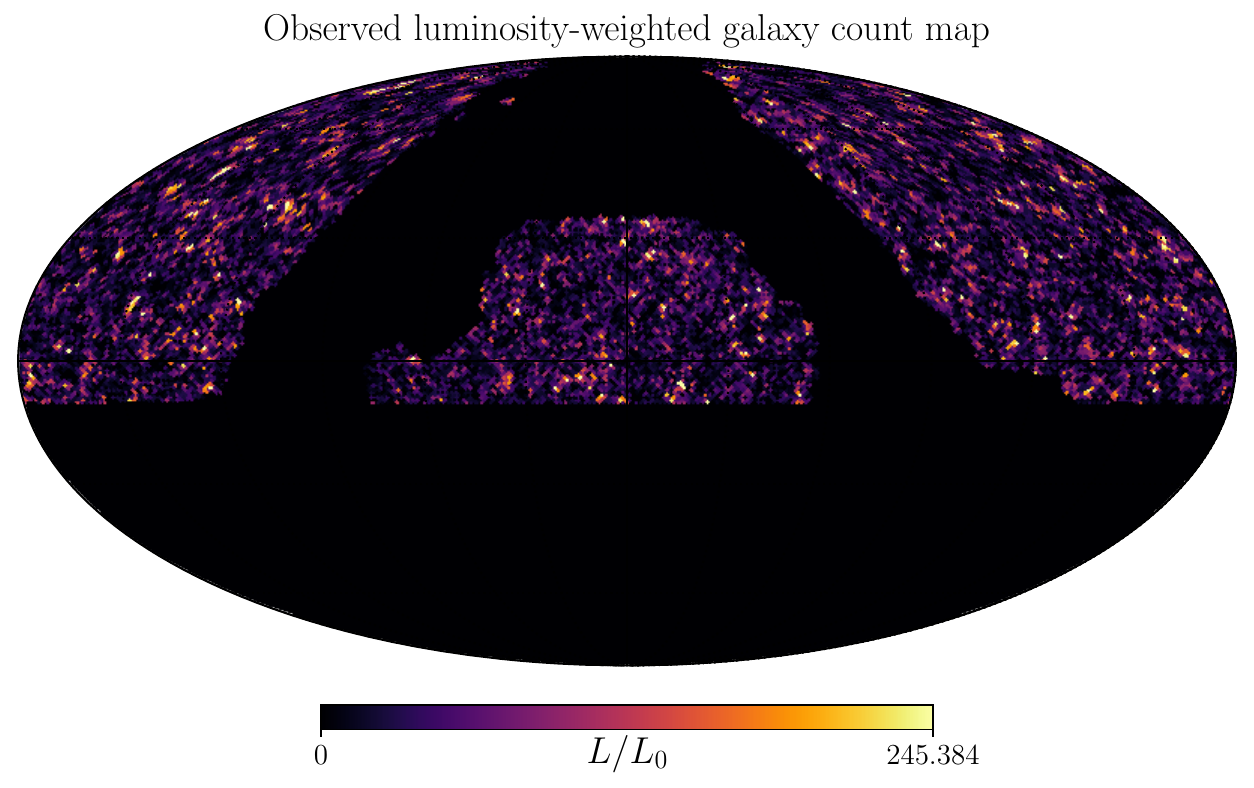}
    \caption{We show here a luminosity-weighted galaxy map at redshift $z = 0.20$. The mask follows the DESI legacy
    footprint with a declination cut $\delta > -10^\circ$. We take this as a proxy for a DESI DR3-like footprint.  
    }
    \label{fig:LW_count_map}
\end{figure}

In practice, galaxy surveys cover only a partial sky, for which spherical harmonics orthogonality breaks. We use the catalogue based $C_\ell$ estimator from {\tt namaster} \cite{Wolz:2024dro} with the DESI weights to recover the expectation value of the auto- and cross-spectra from the pseudo $\t{C}_\ell$'s, which are computed from the masked number counts, via the mode coupling matrix $M_{\ell \ell'}$
\begin{align}
\langle C_\ell^{IJ} \rangle = \sum_{\ell'} M_{\ell \ell'} \t{C}_{\ell'}^{IJ}\,,
\end{align}
and $I,J$ denote either $\rm{g}$ or $\rm{gw}$. The mode-coupling matrix depends on the angular power spectrum of the mask, which is fixed to the survey footprint, for which we show one observed map with $\beta = 1$ in Fig.\,\ref{fig:LW_count_map}. 

\subsection{Peak Sirens simulations}\label{sec:Peak_Sirens_simulation}

We use the same Peak Sirens pipeline described in \cite{Ferri:2024amc} (see also \cite{SantiagodeMatos:2025iyj}) to extract cosmological information from cross-correlations of galaxies and GWs, with the upgrade of accounting for luminosity weighting. This pipeline consists on generating thousands of galaxy and BBH sky maps, in tomographic shells, where the BBH true positions are sampled from the galaxy luminosity weighted distribution. It then allows to estimate the average of the angular power spectra of these maps ($C_{\ell}^{\rm th}$), and its covariance needed for evaluating Eq.\,\eqref{eq:covariance}.

We use the bright galaxy sample of DESI-DR1 \cite{DESI:2025fxa} to estimate the luminosity-dependent bias and the BBH host probabilities. In a fiducial cosmology, we build light-cone simulations using \texttt{GLASS} \cite{Tessore:2023zyk}, where galaxies have only positions, and no other property. We sample from the same matter field realization different galaxy samples with different biases, corresponding to the amount of galaxies in each luminosity bin in DESI, and their respective biases. This is how we inject luminosity bin information into the simulations. We sample the hosts of the BBHs from the luminosity-weighted maps, which we then mask according to a DESI DR3-like footprint. We build the auto- and cross- spectra from the masked skies using {\tt namaster} \cite{Alonso:2018jzx} for the fiducial cosmology. We then use the mean of these simulations to estimate the theoretical $C_\ell$'s and covariance using Eq.\,\eqref{eq:covariance} and apply a rescaling technique pioneered by \cite{Ferri:2024amc} to evaluate the theoretical $C_\ell$'s and covariance in different cosmologies. 

The sky map of each BBH is further deformed into a pixelized probability density map, according to the expected uncertainties in the distance and sky position of the future network of GW detectors. For this, we use the same GW Fisher forecasts as in \cite{Ferri:2024amc}, generated with the code GWDALI \cite{deSouza:2023ozp}. These are catalogs containing the expected precision on 11 parameters specifying each GW source: masses, spins (neglecting precession), distance, inclination angle of the binary's angular momentum w.r.t the line-of-sight, polarization angle, time and phase of coalescence, and sky position angles. We consider the Einstein Telescope in combination with two Cosmic Explorer (ET+2CE) \cite{ET:2019dnz,Reitze:2019iox,Kalogera:2021bya}. For the ET+2CE configuration where ET is assumed in its triangular shape, our conservative estimate results in around 70,000 detectable sources in 5 years with 80\% duty cycle, after the cut SNR$>$12, and after excluding low inclination binaries, where Fisher forecasts are problematic \cite{deSouza:2025qok}. This assumes a merger rate model based on the star-formation rate, convolved with a log-uniform time delay. The value of the merger rate today is taken to be in agreement with the latest GW observations, noticing it has large uncertainty up to this date \cite{LIGOScientific:2026ctl}. We also assume a Power law + Peak mass-model. Whereas the precise number of sources and precision of parameter estimation depend on the mass model, it should not considerably affect the conclusions from the cross-correlations method obtained in this work \cite{Cheng:2026atn}.

\section{Results}\label{sec:results}
At this stage, we forecast the abilities of a DESI-like survey of bright galaxies to probe cosmology from the cross-correlations with GW sources, employing different luminosity-weighted strategies. For this purpose, we focus on the abilities of ET+2CE to measure the parameters $\{H_0, \Omega\e{m}, b\e{gw}\}$. We show in particular that a measurement of the GW linear bias better than 10\% accuracy is expected with ET+2CE and a DESI-like Bright Galaxy survey and that tight measurements of $H_0$ and $\Omega\e{m}$ of $0.05\%$ and $5\%$ are robust to a mismodelling of the GW bias.

\subsection{Luminosity weighting}

In this subsection, we apply different weights to the luminosity-weighted galaxies, via Eq.\,\eqref{eq:weighted_counts}. We weight galaxies with the same prescription both when sampling the GW host galaxies and when building the weighted galaxy map with which we cross-correlate the GW maps. In particular, we compare linear ($\beta = 1$) and quadratic ($\beta = 2$) with uniform weighting ($\beta = 0$). Because of the relatively small bias difference between the different luminosity bins depicted in Fig.\,\ref{fig:Effective_bias}, the constraints are robust to changes in the luminosity weights, as may be seen from the contour plots presented in Fig.\ref{fig:Contour_Lweight=0,1,2_et2ce}. Note that this contrasts with the dark siren method where the authors of \cite{Perna:2024lod} have shown that mismodelling the luminosity weight may generate a bias in the cosmological inference. 
    \begin{figure}
    \includegraphics[width=1.0\linewidth]{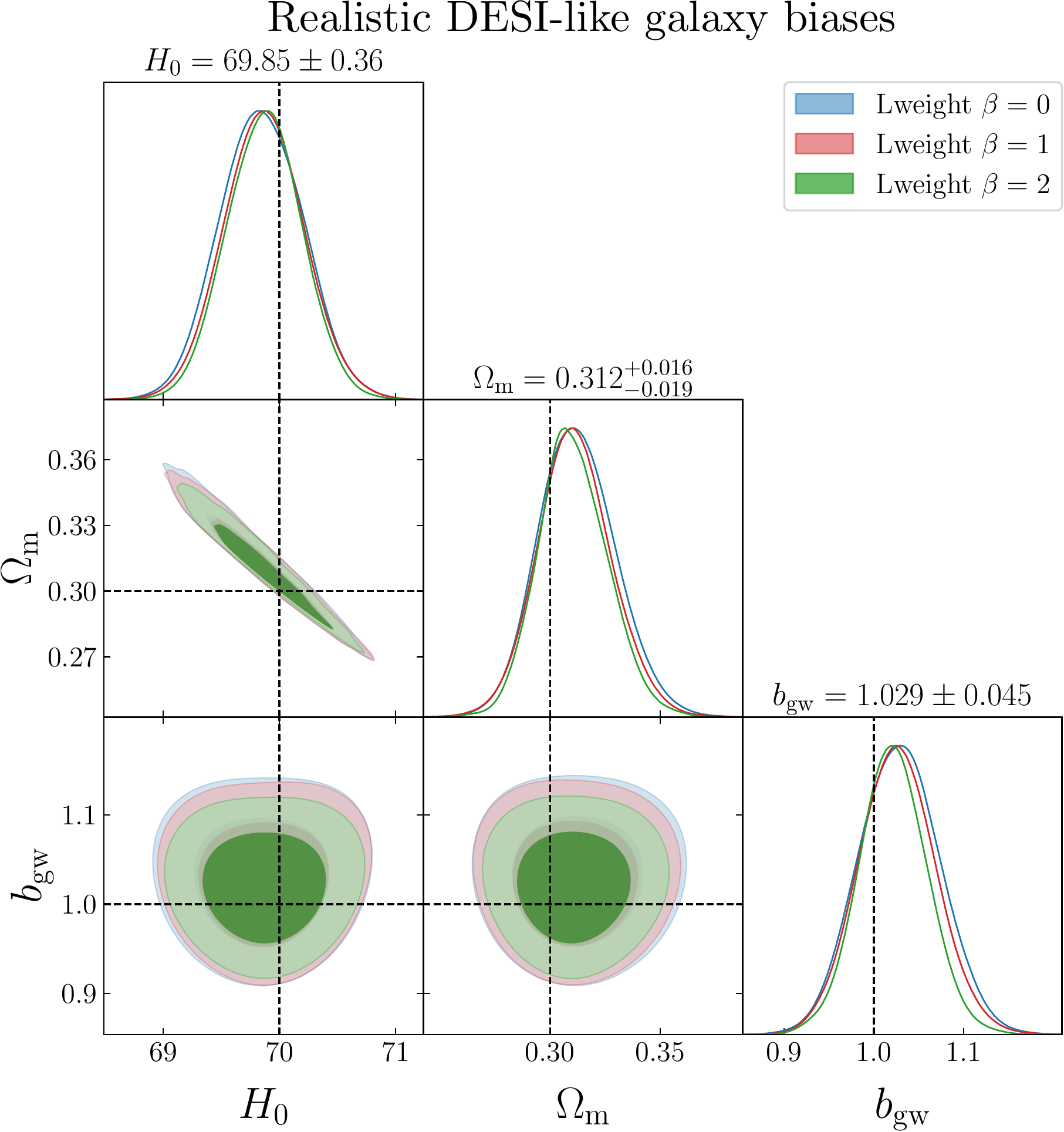}
    \caption{For $\ell\e{max} = 190$, we present the contour plots for ET+2CE with redshift bin width $\Delta z = 0.01$ between $z\in [0,0.5]$. We present the contours for uniform ($\beta =0$), linear ($\beta = 1$) and quadratic ($\beta = 2$) weighting in blue, red and green, respectively. The contours overlap, although there is a small tightning of the posteriors with higher $\beta$. The marginalized uncertainties on top of each plot are those of the uniformly weighted ($\beta =0$) case. The dashed lines show the fiducial model. }
    \label{fig:Contour_Lweight=0,1,2_et2ce}
\end{figure}
The changes in the figures of merit follow from the change in the effective galaxy bias, presented in Fig.\,\ref{fig:Effective_bias} which affects the GW bias and the SNR via Eq.\,\eqref{eq:SNR_bgw}. One way to generate a more significant change in the effective bias is if the brightest galaxies are more clustered than their fainter counterpart. Next, we explore a toy model, where the galaxies from the brightest luminosity bin have a bias 4 times larger than the other galaxies. In this situation, the contours can shrink more significantly, as presented in Fig.\,\ref{fig:Contour_Lweight=0TM,1TM,2TM_et2ce}. 

\begin{figure}
    \includegraphics[width=0.9\linewidth]{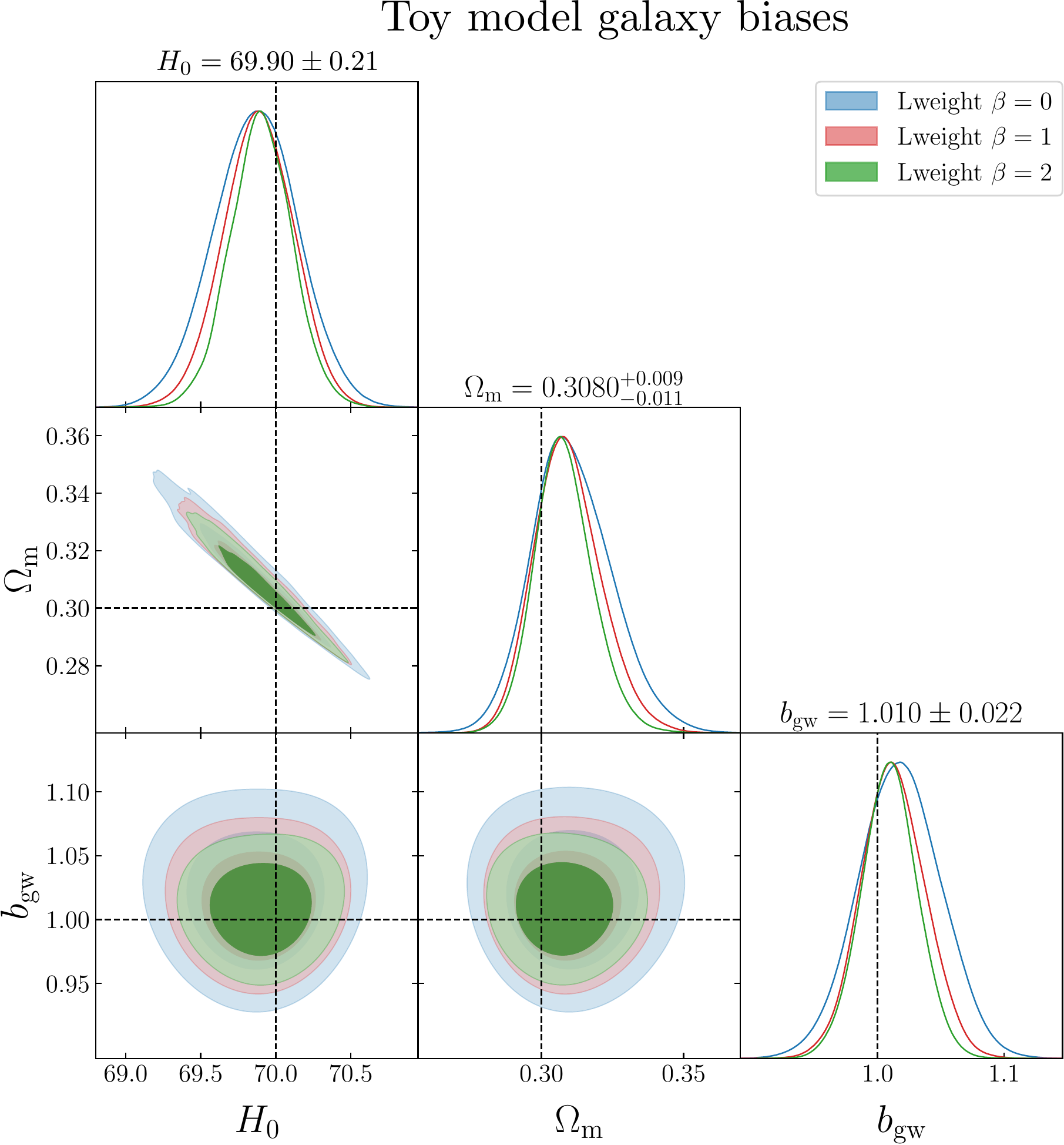}
    \caption{For $\ell\e{max} = 190$, we present the contour plots for ET+2CE with redshift bin width $\Delta z = 0.01$ between $z\in [0,0.5]$. We present the contours for $\beta =0$, $\beta =1$ and $\beta =2$ in blue, red and green, respectively. The toy model galaxy biases are $\bs{b}\e{g} =(1,1,1,1,4)$ for the 5 different luminosity bins. 
    In this case, increasing the weighting of luminous galaxies tightens the error bars.
    }
    \label{fig:Contour_Lweight=0TM,1TM,2TM_et2ce}
\end{figure}

It was noted in \cite{Perna:2024lod} that fixing the luminosity weighting model within the line of sight method may introduce a bias in cosmological parameter. Along these lines, the authors of \cite{Li:2025hrh} proposed a Bayesian method to recover the preferred model among a set of luminosity weighting prescriptions. Instead, the authors of \cite{Agapito:2026pbb} proposed to infer the luminosity weights jointly with cosmology. Our results suggest that the cosmological inference of the Peak Sirens method is weakly sensitive to the luminosity weighting prescription. In the next section, we show that the method is robust to a mismodelling of the GW bias.

\subsection{Gravitational wave bias.}
\begin{figure}
    \includegraphics[width=1.0\linewidth]{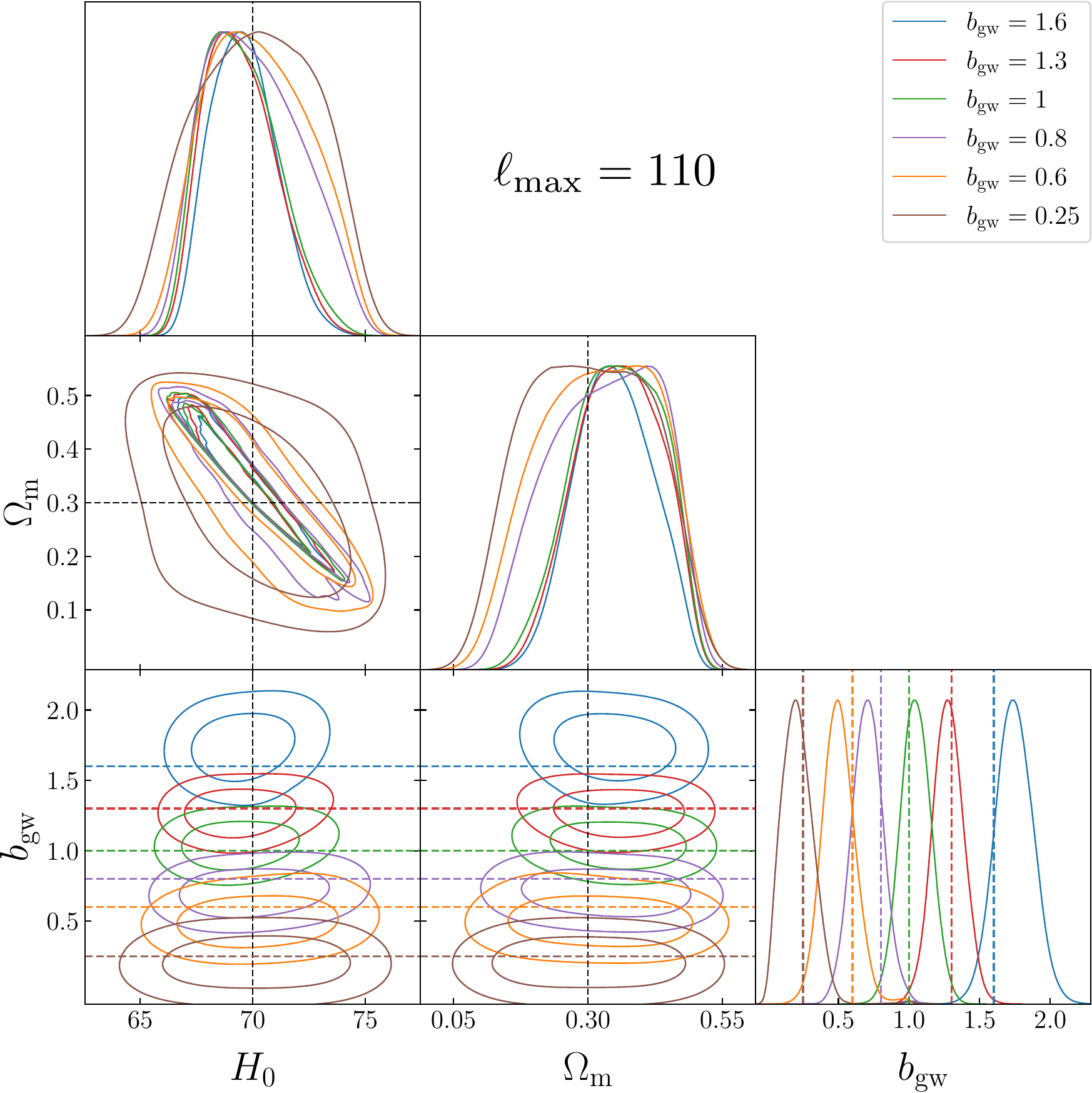}
    \caption{For $\ell\e{max} = 110$, we present the contour plots for ET+2CE where the observed GWs are sampled from a population with bias $b\e{gw} \in \{ 0.25,0.6, 0.8, 1,1.3,1.6\}$. The observer performs the simulations assuming $b\e{gw} = 1$ in all cases and recovers the injected GW bias in all cases. The uncertainties in $H_0$ and $\Omega_m$ shrink with increasing $b\e{gw}$, as expected from Eq.\,\eqref{eq:SNR_bgw}. To recover the linear bias requires to restrict to relatively large scales. 
    }
    \label{fig:Contour_bgw_et2ce_lmax=110}
\end{figure}

\begin{center}
\begin{figure*}
    \includegraphics[width=1.0\linewidth]{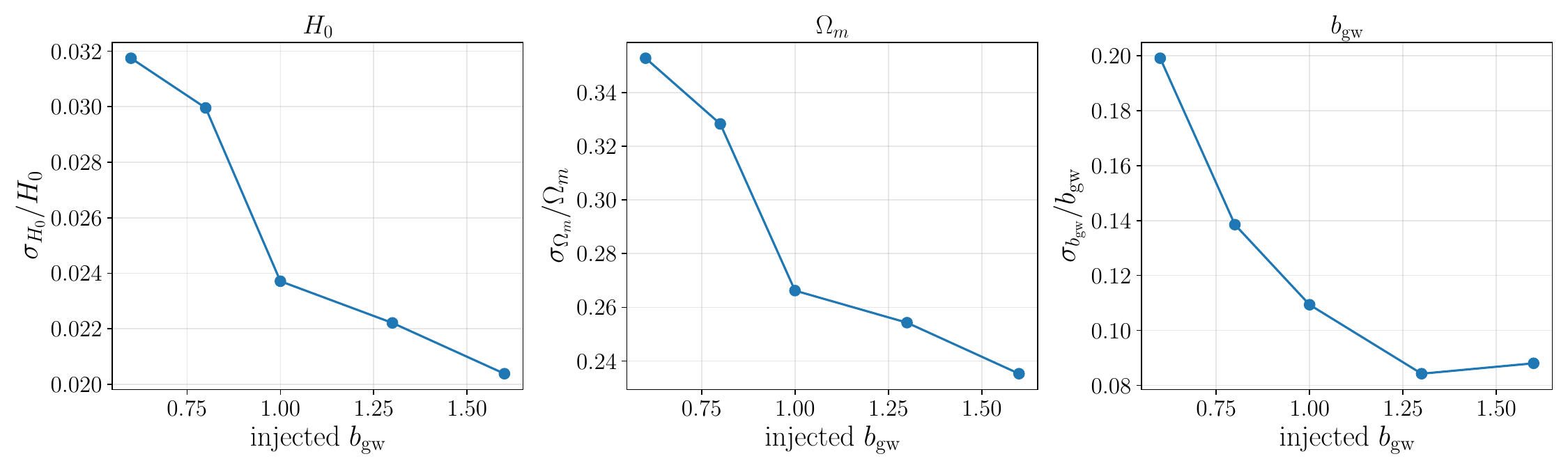}
    \caption{For $\ell\e{max} = 110$, we present the relative uncertainties on $\{H_0, \Omega\e{m}, b\e{gw}\}$ as a function of the bias injected for the host galaxies of GWs. The simulations are performed with galaxies that have DESI-like linear biases, as presented in Fig.\,\ref{fig:bias_parameter_L_NLbins=5}, while the GW are simulated such that $b\e{gw} = 1$. The observed $C_\ell$'s are taken from simulations with $b\e{gw} \in \{0.25,0.6,0.8,1,1.3,1.6\}$ to check how well one can reconstruct the cosmological parameters and the GW bias. The relative uncertainties drop as $b\e{gw}$ gets higher, as expected from the SNR calculation, presented in Eq.\,\eqref{eq:SNR_bgw}. }
    \label{fig:sigmaH0_bgw}
\end{figure*}
\end{center}

In the previous sections, we noted that raising the effective bias of host galaxies raises the signal of the cross-correlation in a way which tightens the constraints on cosmological parameters and the GW bias. This immediately raises the following question: What if the GWs come from a subpopulation of galaxies with a different bias than the galaxies that are seen ? In the following, we perform the cross-correlation with the observed galaxies but sample the host galaxies from another population of galaxies which are not seen by the observer and which have a different bias. We show that the methodology is robust to this mismatch and that we are able to recover the injected bias for the GW hosts in Fig.\,\ref{fig:Contour_bgw_et2ce_lmax=110}. There, it is required to stick to relatively large scale $\ell\e{max}=110$ to recover the injected GW linear bias. Indeed, working with $\ell\e{max} = 190$ may lead to a bias on the GW bias, although cosmological parameters are recovered without any bias, as discussed in the next subsection. This probably comes from nonlinearities and the breaking of the scale independence of the GW linear bias. 
We show how the uncertainties on $H_0$, $\Omega\e{m}$ and $b\e{gw}$ depend on $b\e{gw}$ in Fig.\,\ref{fig:sigmaH0_bgw}.

\subsection{Robustness of cosmological parameter constraints}

\begin{figure}
    \includegraphics[width=1.0\linewidth]{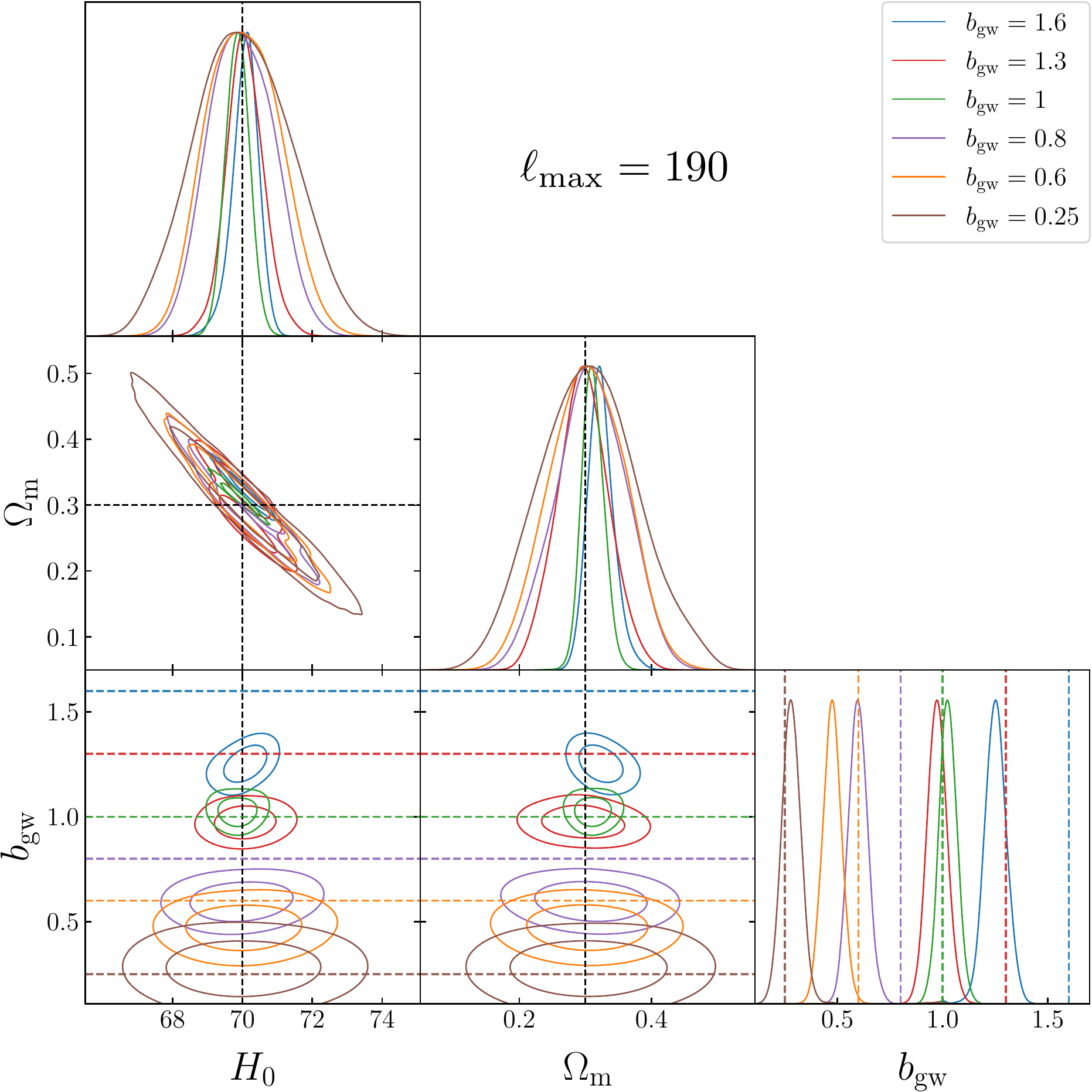}
    \caption{For $\ell\e{max} = 190$, we present the contour plots for ET+2CE where the observed GWs are sampled from a population with bias $b\e{gw} \in \{ 0.25,0.6, 0.8, 1,1.3,1.6\}$. The observer performs the simulations assuming $b\e{gw} = 1$ in all cases and only recovers the injected bias in the subcases $b\e{gw} \in\{0.25, 1\}$. Nonetheless, $H_0$ and $\Omega\e{m}$ are recovered within $1\sigma$. To recover robustly the linear bias of GWs requires to restrict to larger scales with $\ell\e{max} = 110$, as presented in Fig.\,\ref{fig:Contour_bgw_et2ce_lmax=110}. }
    \label{fig:Contour_bgw_et2ce_lmax=190}
\end{figure}

In the previous subsection, we showed that the GW linear bias can be recovered even if the observer misses the true host galaxies which have a different linear bias than the galaxies that are observed. To achieve this result, the maximal resolution was set to $\ell\e{max}=110$, avoiding the effect of nonlinear scales on the linear bias. Interestingly, pushing to higher $\ell\e{max} = 190$ which is set be the resolution of the simulations generates a bias on the GW bias while lowering robustly the uncertainties on the Hubble constant and matter energy density. We present the contours for $\ell\e{max} = 190$ in Fig.\,\ref{fig:Contour_bgw_et2ce_lmax=190}. We find that with $\ell\e{max} = 190$ and $b\e{gw}=1$, one can robustly constrain $H_0$ to $0.5\%$ and $\Omega\e{m}$ to $5\%$, despite the mismodelling of the GW bias. We summarize the constraints on $\{H_0,\Omega\e{m}, b\e{gw}\}$ for $b\e{gw}=1$ depending on $\ell\e{max}$ in Tab.\ref{tab:constraints}.
\begin{table}[htbp] 
\renewcommand{\arraystretch}{1.5}
\begin{tabular}{l|ccc} 
$\ell\e{max}$ & $\sigma_{H_0}/H_0 $& $\sigma_{\Omega\e{m}}/\Omega\e{m}$ & $\sigma_{b\e{gw}}/b\e{gw}$ \\ \hline 
110 & 2\%& 26\% & 10\% \\
190 & 0.5\% & 5\% & N/A
\end{tabular} 
\caption{We present our forecast on $\{H_0,\Omega\e{m}, b\e{gw}\}$ for two relevant values of the maximal resolution $\ell\e{max}$. Using $\ell\e{max}= 110$ allows to robustly infer the GW linear bias, even if the GWs come from a population of unobserved galaxies which have a different bias than those that are observed. If one pushes the resolution to $\ell\e{max} = 190$, the GW linear bias becomes biased, probably due to nonlinear effects which make the GW bias a function of redshift and scale. However, quite encouraging is the fact that the constraints on $H_0$ and $\Omega\e{m}$ are consistent with the injected values within $1\sigma$ with tighter error bars than with $\ell\e{max} = 110$. }\label{tab:constraints}
\end{table}

\begin{figure*}
    \includegraphics[width=0.7\linewidth]{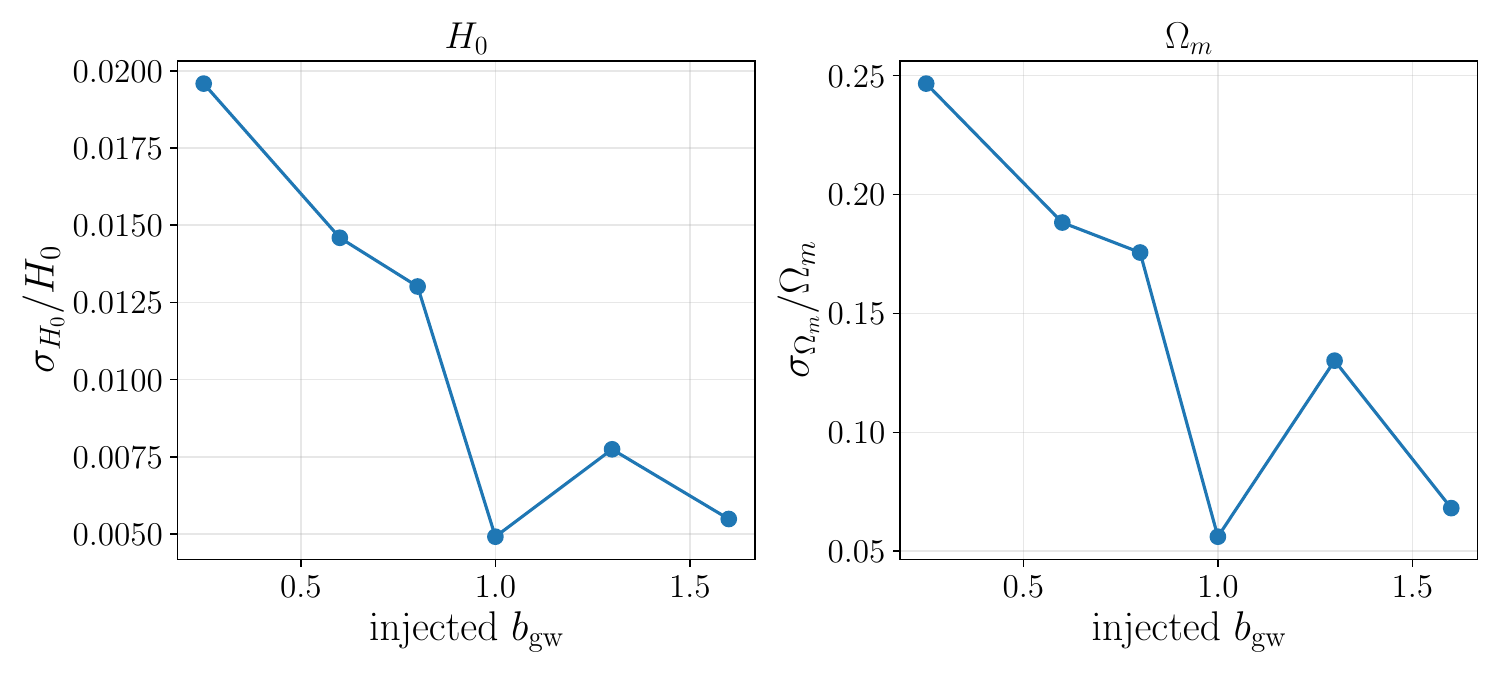}
    \caption{For $\ell\e{max} = 190$, we present the relative uncertainties on $\{H_0, \Omega\e{m}\}$ as a function of the bias injected for the host galaxies of GWs. The uncertainties are tighter than for $\ell\e{max} = 110$, as expected. The simulations are performed with galaxies that have DESI-like linear biases, as presented in Fig.\,\ref{fig:bias_parameter_L_NLbins=5}, while the GW are simulated such that $b\e{gw} = 1$. The observed $C_\ell$'s are taken from simulations with $b\e{gw} \in \{0.25,0.6,0.8,1,1.3,1.6\}$ to check how well one can reconstruct the cosmological parameters. The relative uncertainties drop as $b\e{gw}$ gets higher, as expected from the SNR calculation, presented in Eq.\,\eqref{eq:SNR_bgw}, although there is a dip in the uncertainties for an injected $b\e{gw}$ which matches the one in the fiducial model ($b\e{gw}=1$). The bias of GWs becomes biased at these scales, as depicted in Fig.\,\ref{fig:Contour_bgw_et2ce_lmax=190}.}
    \label{fig:sigmaH0_bgw_lmax=190}
\end{figure*}

\begin{figure}
    \centering
    \includegraphics[width=1.0\linewidth]{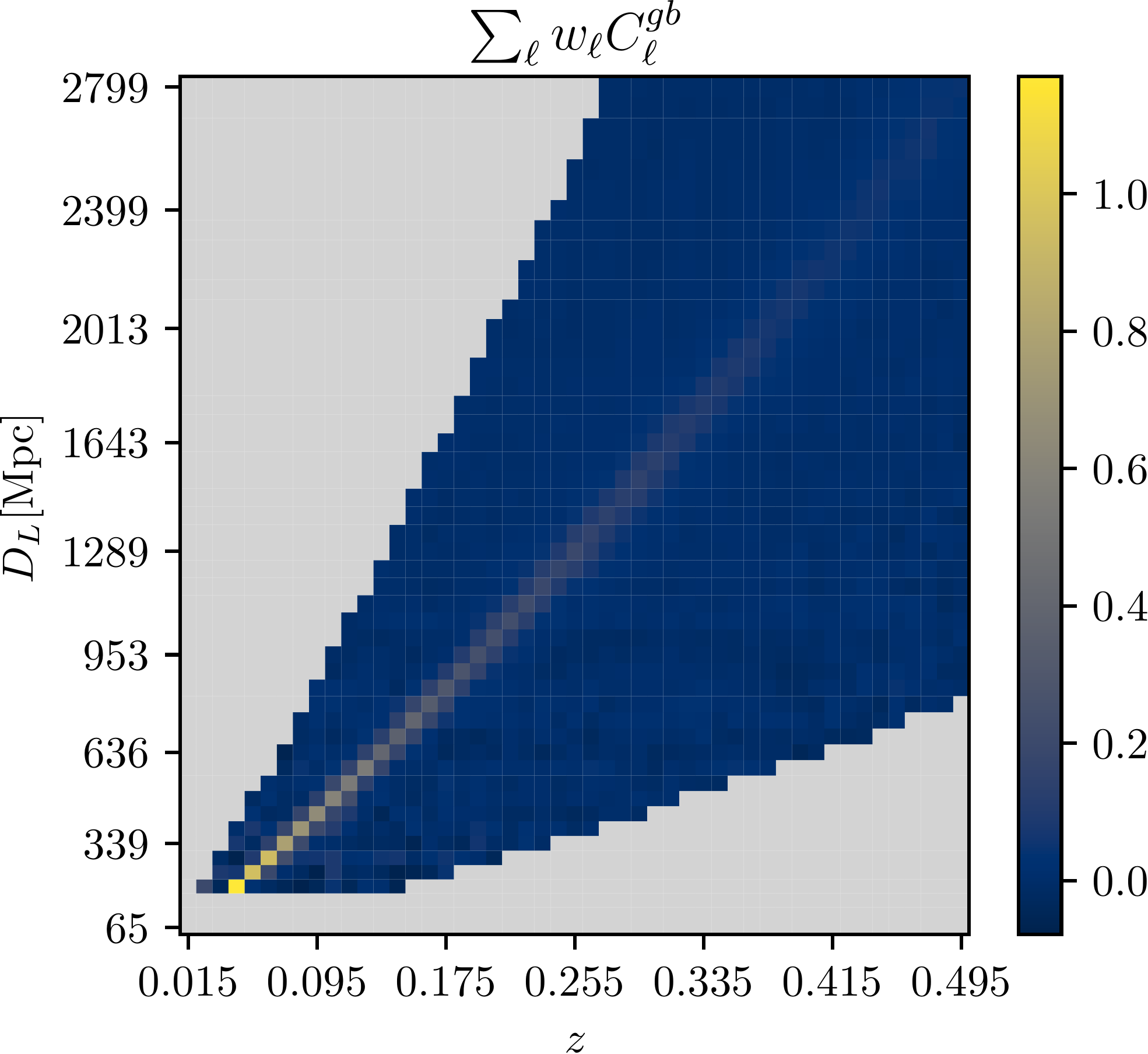}
    \caption{Here we plot the average $C_\ell^{gb}(z, D_L)$ for 1000 simulations with ET2CE setting. The cross-correlation signal is clearly visible along the diagonal. The lower distance bins are cut as they host no GWs and the monopole vanishes making the GW density contrast ill-defined. 
    }
    \label{fig:CL_gb_mean_Lweight=0}
\end{figure}

\section{Conclusion}\label{sec:Conclusion}

In this work, we investigated the role of luminosity weighting in the cross-correlations of galaxies and gravitational waves. We presented a formalism, where galaxies are split in different luminosity bins with a different bias prescription for each population. We used DESI DR1 data to obtain realistic input for these biases and luminosities and found that higher luminosity objects indeed have a higher linear bias. For these realistic values, the change in the effective bias of the host galaxies is relatively modest. We ran simulations with different bias prescriptions for each population and found that the constraints on the cosmological parameters are robust to these changes. In contrast, if there is a more important gradient in the effective bias as a function of luminosity weight, then the constraints on cosmological parameters can improve, although with an extra unknown parameter, which is the luminosity weight. This may open the possibility to measure the luminosity weight, although we leave this possibility for future work. We illustrated this effect with a toy model in which the brightest galaxies have a linear bias 4 times larger than the other galaxies, which resulted in a shrinkage of the contours. The weak sensitivity of the method to luminosity weighting means that the method is robust to different luminosity weighting prescriptions, that is, to the modeling of how likely galaxies are to host GW sources.

We then investigated what happens if the observer misses the true host galaxies such that those have a different bias than the galaxies that are observed. We showed how the GW bias can be recovered, although this required to focus the analysis on large scale with $\ell\e{max}\leq 110$. We obtained robust constraints on the GW bias of the order of $10\%$ for $b\e{gw}=1$. We showed how the uncertainties on $H_0$, $\Omega\e{m}$ and $b\e{gw}$ depend on the GW bias, finding that they shrink as one increases the GW bias, as expected. We further showed that one may use smaller scales up to the maximal resolution of the simulations (which is $\ell\e{max}=190$) to further constrain the Hubble constant and matter energy density. We found that despite the GW bias being contaminated by nonlinearities, $H_0$ and $\Omega\e{m}$ can robustly be determined to $0.5\%$ and $5\%$ precision, respectively for $b\e{gw}=1$. We conclude that cross-correlations are robust to different luminosity weighting prescriptions of galaxies and to a mismodelling of the GW bias in the simulations. Whereas the host probability is a key ingredient of the dark sirens method which in most available pipelines has to be modeled, we showed that for cross-correlations this modeling is not important. This is encouraging for the method to be applied to more complete and extensive galaxy catalogues such as DESI.

\noindent{\bf Acknowledgements} 

Numerical computations were carried out on the Thanos High Performance Computer (HPC) cluster which is supported by the ENS. I.S.M.\, and T.B.\,are supported by ERC Starting Grant SHADE (Grant No.\,StG 949572). T.B.\,is further supported by a Royal Society University Research Fellowship (Grant No.\,URF\textbackslash R\textbackslash 231006). We acknowledge use of \texttt{Healpy} \cite{Zonca2019, 2005ApJ...622..759G}, \texttt{Matplotlib} \cite{Hunter:2007}, \texttt{astropy} \cite{astropy:2022}. This research uses data from the Dark Energy Spectroscopic Instrument (DESI) Data Release 1 (DR1) \cite{DESI:2025fxa}, accessed through CosmoHub \cite{Carretero:2017zkw}.

\bibliographystyle{apsrev4-2}
\bibliography{references} 

\end{document}

%% file: macros.tex
\def \be {\begin{equation}}
\def \ee {\end{equation}}
\def \dd {\mathrm{d}} 
\def \t {\tilde}

\def \l {\left}
\def \r {\right}

\def \bs {\boldsymbol}

\newcommand{\e}[1]{_{\rm #1}}

\newcommand{\beq}{\begin{equation}}
\newcommand{\eeq}{\end{equation}}
\newcommand{\bea}{\begin{eqnarray}}
\newcommand{\eea}{\end{eqnarray}}


\newcommand\ees{\end{eqnarray}}
\newcommand\bees{\begin{eqnarray}}

\def\dd{\mathrm{d}}
